\RequirePackage{fix-cm} 
\RequirePackage{fixltx2e}
\RequirePackage[reqno]{amsmath}
\RequirePackage[l2tabu,orthodox]{nag}
\documentclass[final,dvips, a4paper, twoside, reqno, 11pt]{amsart}
\usepackage{etex}

\usepackage[T1]{fontenc}
\usepackage[latin1]{inputenc}
\usepackage{lmodern}

\usepackage{esint}
\usepackage{amsbsy}
\usepackage{dsfont}
\usepackage{empheq}
\usepackage{amsthm}
\usepackage{manfnt}
\usepackage{pifont}
\usepackage{xspace}
\usepackage{amssymb}
\usepackage{siunitx}
\usepackage{upgreek}
\usepackage{amsfonts}
\usepackage{extarrows}
\usepackage{mathtools}
\usepackage{textgreek}

\usepackage{mathabx}
\usepackage{textcomp}

\usepackage{mathrsfs}
\DeclareMathAlphabet{\mathscrbf}{OMS}{mdugm}{b}{n}

\DeclareFontFamily{U}{dutchcal}{\skewchar\font=45 }
\DeclareFontShape{U}{dutchcal}{m}{n}{<-> s*[1.0] dutchcal-r}{}
\DeclareFontShape{U}{dutchcal}{b}{n}{<-> s*[1.0] dutchcal-b}{}
\DeclareMathAlphabet{\mathlcal}{U}{dutchcal}{m}{n}
\SetMathAlphabet{\mathlcal}{bold}{U}{dutchcal}{b}{n}

\usepackage{a4wide}
\usepackage[square,numbers]{natbib}

\usepackage[dvipsnames]{xcolor}

\definecolor{bckg}{RGB}{20.8, 20.8, 20.8}
\definecolor{oneblue}{rgb}{0.0, 0.0, 0.85}
\definecolor{Lightblue}{RGB}{214, 214, 214}
\definecolor{bluepigment}{rgb}{0.2, 0.2, 0.6}
\definecolor{charcoal}{rgb}{0.21, 0.27, 0.31}
\definecolor{denimblue}{rgb}{0.08, 0.38, 0.74}
\definecolor{darkelectricblue}{rgb}{0.33, 0.41, 0.47}
\definecolor{katyblue}{rgb}{0.129412, 0.137255, 0.63}

\usepackage{booktabs}

\usepackage{graphicx}
\graphicspath{{figs/}}
\DeclareGraphicsExtensions{.pdf,.eps}

\usepackage{tikz}
\usepackage{subfigure}
\usepackage{morefloats}
\usepackage{indentfirst}

\usepackage[labelsep=period,
            font=normalsize,
            labelformat=simple,
            labelfont={bf,sf,color=bluepigment},
            justification=raggedright]{caption}

\usepackage[perpage, multiple, symbol]{footmisc}
\setfnsymbol{wiley}

\newcommand*{\Title}{Regularised shallow water equations with uneven bottom}
\newcommand*{\Longtitle}{Hamiltonian regularisation of shallow water equations with 
uneven bottom}

\newcommand*{\Keywords}{shallow water flow; dispersion; regularisation; energy 
conservation; well-balanced; uneven bottom}

\usepackage{url}
\usepackage[
           final=true,
           linktoc=all,
           unicode=true,
           bookmarks=true,
           colorlinks=true,
           breaklinks=true,
           urlcolor=katyblue,
           linkcolor=MidnightBlue,
           citecolor=NavyBlue,
           pdffitwindow=false,
           bookmarksopen=false,
           pdfpagemode=UseNone,
           plainpages=false,
           hyperfigures=true,
           pdfstartview={FitV},
           pdftitle={\Longtitle},
           pdfkeywords={\Keywords},
           pdfnewwindow=true,
           backref=true,
           pagebackref=true]{hyperref}

\usepackage[sort&compress, capitalise, noabbrev]{cleveref}
\usepackage[all]{hypcap}

\usepackage[square,numbers]{natbib}
\usepackage{enumitem}
\setlist[description]{%
  topsep = 30pt,               
  itemsep = 8pt,               
  labelsep = 10pt,
  font={\bfseries\color{NavyBlue}}, 
}

\numberwithin{equation}{section}

\theoremstyle{remark}

\newcommand{\ud}{\mathrm{d}}

\renewcommand{\eta}{\hspace{0.05em}\mbox{\texteta}}

\newcommand{\dbar}{d\hspace*{-0.08em}\bar{}\hspace*{0.1em}}

\newcommand{\eps}{\epsilon}

\newcommand{\eqdef}{\mathop{\stackrel{\,\mathrm{def}}{=}\,}}

\newcommand{\scal}{\;\raisebox{0.25ex}{\tikz\filldraw[black, x=1pt, y=1pt](0,0) circle (1);}\;}

\newcommand{\half}{{\textstyle{1\over2}}}
\newcommand{\sixth}{{\textstyle{1\over6}}}

\newcommand{\threehalf}{{\textstyle{3\over2}}}

\begin{document}

\title[\Title]{\Longtitle}

\author[D.~Clamond]{Didier Clamond$^*$}
\address{{\bf D.~Clamond:} Universit\'e C\^ote d'Azur, CNRS UMR 7351, LJAD, Parc Valrose, F-06108 Nice cedex 2, France}
\email{\href{mailto:didier.clamond@unice.fr}{didier.clamond@unice.fr}}
\urladdr{\url{http://math.unice.fr/~didierc/}}
\thanks{{$^*$}\it Corresponding author}

\author[D.~Dutykh]{Denys Dutykh}
\address{{\bf D.~Dutykh:} Univ. Grenoble Alpes, Univ. Savoie Mont Blanc, CNRS, LAMA, 73000 Chamb\'ery, France and LAMA, UMR 5127 CNRS, Universit\'e Savoie Mont Blanc, Campus Scientifique, 73376 Le Bourget-du-Lac Cedex, France}
\email{\href{mailto:Denys.Dutykh@univ-smb.fr}{Denys.Dutykh@univ-smb.fr}}
\urladdr{\url{http://www.denys-dutykh.com/}}

\author[D.~Mitsotakis]{Dimitrios Mitsotakis}
\address{{\bf D.~Mitsotakis:} Victoria University of Wellington, School of Mathematics, Statistics and Operations Research, PO Box 600, Wellington 6140, New Zealand}
\email{\href{mailto:dmitsot@gmail.com}{dmitsot@gmail.com}}
\urladdr{\url{http://dmitsot.googlepages.com/}}



\maketitle


\begin{abstract}

The regularisation of nonlinear hyperbolic conservation laws has been a problem  of great importance 
for achieving uniqueness of weak solutions and also for accurate numerical simulations. In a recent 
work, the first two authors proposed a so-called Hamiltonian regularisation for nonlinear shallow 
water and isentropic Euler equations. The characteristic property of this method is that the 
regularisation of solutions is achieved without adding any artificial dissipation or dispersion. 
The regularised system possesses a Hamiltonian structure and, thus, formally preserves the corresponding 
energy functional. 
In the present article we generalise this approach to shallow water waves over general, possibly 
time-dependent, bottoms. The proposed system is solved numerically with continuous Galerkin method and 
its solutions are compared with the analogous solutions of the classical shallow water and dispersive 
Serre--Green--Naghdi equations. The numerical results confirm the absence of dispersive and dissipative 
effects in presence of bathymetry variations.


\end{abstract}


\section{Introduction}

Many phenomena in fluid mechanics are described mathematically by systems of hyperbolic 
equations \citep{Lions1998}. We can mention the celebrated inviscid Burgers--Hopf equation 
\citep{Burgers1948} as a prototype of pressureless Euler equations, the isentropic Euler 
equations \citep{MajdaBertozzi2001}, the shallow water (Airy or Saint-Venant) equations 
\citep{SaintVenant1871}, the compressible Euler equations \citep{Lions1998} and even some 
two-phase flow models \citep{DiasEtAl2010}. These equations have a common property: if we 
solve an initial value problem with infinitely smooth (or even analytic) data, the solutions 
will develop a finite time singularity (e.g., a gradient ``catastrophe''). Thus, to speak 
mathematically about these solutions, one has to introduce the so-called {\em weak solutions} 
\citep{DiPernaMajda1987}, or even weaker than weak solutions \citep{Hong2006}. One strategy 
employed by mathematicians to study such systems consists in considering a perturbed version 
of equations with a perturbation being chosen so that the new (perturbed) system has more 
regular (i.e., smoother) solutions. The original system of governing equations can be formally 
recovered as a singular limit of the perturbed system. Then, some conclusions about weak solutions 
of the original system are obtained by employing the bootstrap argument \citep{Lions1998}. The 
perturbation is usually chosen to be of dissipative, dispersive or of both types 
\citep{LeFlochNatalini1999}. 
We can mention a few previous attempts to regularise the inviscid Burgers--Hopf equation with 
dissipative/dispersive terms \citep{BonaSchonbek1985,BrenierLevy2000}.

In a recent work, \citet{ClamondDutykh2018a} propose a regularisation of the nonlinear shallow water 
(or Saint-Venant) equations (NSWE) with flat bottom, that describe long gravity waves propagating 
in both directions under the hydrostatic pressure assumption. In particular, these regularised 
Saint-Venant (rSV) equations are a conservative Hamiltonian system that regularises the solutions 
of the NSWE without adding any artificial dissipation or dispersion. Some properties of these regularised 
shallow water equations are mathematical study in \citep{LiuEtAl2019,PuEtAl2018}. 

The goal of the present manuscript is to generalise the approach proposed by \citet{ClamondDutykh2018a} 
to general uneven and time-dependent bottoms. The latter might be useful for tsunami-generation problems 
\citep{DutykhDias2009}. The model we derive below conserves all the good properties of regularised 
Saint-Venant equations (such as the energy conservation) despite bathymetry variations in space and in 
time. 

We note that the rSV equations are a two-component generalisation of the dispersionless 
\citet{CamassaHolm1993} (CH) equation. Shallow water equations, such as KdV, KP and CH, 
are also known to play a fundamental role in theoretical Physics and in Geometry 
\citep{HolmEtAl1998,Miller2007,Misiolek1998}. Therefore, the rSV equations may be of 
general physical and mathematical interest. 

The present manuscript is organised as follows. In section \ref{sec:mmodel}, we point out 
the shortcomings of the rSV equations (as proposed in \citep{ClamondDutykh2018a}) for varying 
bottoms and we address  these limitations in order to obtain a suitable regularisation of the 
NSWE for general bottoms. In particular, the Hamiltonian structure of the obtained system 
is also highlighted in this section. The obtained system is briefly studied numerically in 
section \ref{sec:num}, providing numerical evidences that we indeed derived a dispersionless 
Hamiltonian regularisation of the Saint-Venant equations. The main conclusions and perspectives 
of this study are outlined in the section \ref{sec:cp}.


\begin{figure}
  \centering\capstart
  \includegraphics[width=0.59\textwidth]{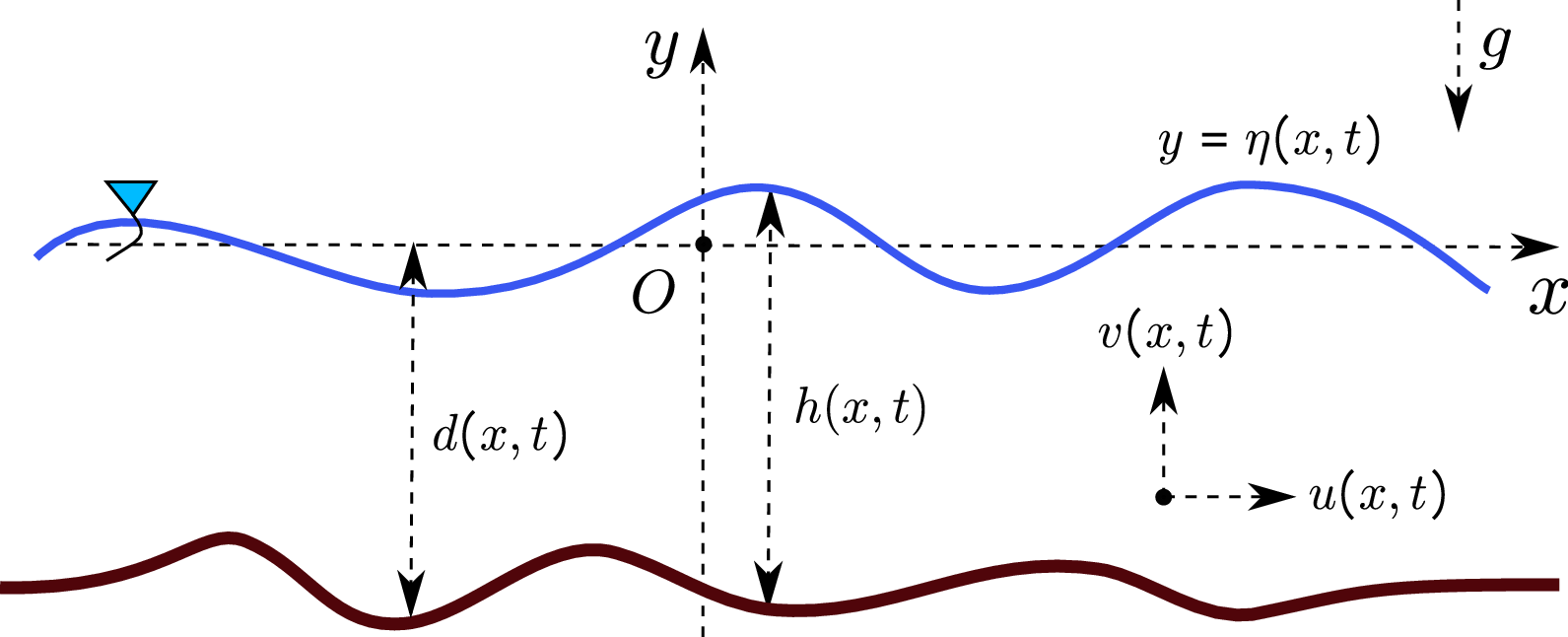}
  \caption{\em Definition sketch.}
  \label{fig:sketch}
\end{figure}
\section{Mathematical model} \label{sec:mmodel}

We consider a two-dimensional irrotational motion due to a gravity wave propagating at the free 
surface of an ideal, incompressible and homogeneous shallow fluid.
Let $x$, $y$ and $t$ be the horizontal, upward vertical and temporal coordinates, respectively. 
The equations $y=0$, $y=\eta(x,t)$ and $y=-d(x,t)$ denote, respectively, the equations of the 
still water level, of the impermeable free surface and of the impermeable bottom; $h\eqdef\eta+d$ 
denoting the total eight of the water column. The parameters $g$ and $\rho$ denote, respectively, 
the acceleration due to gravity directed downwards and the constant fluid density. A sketch of the 
fluid domain is shown in \cref{fig:sketch}.

The definition of the still water level yields
\begin{equation}
  \langle\,\eta\,\rangle\ =\ 0,
\end{equation}
where $\langle\cdot\rangle$ denotes the horizontal Eulerian averaging. The mean water depth is
\begin{equation}
  \dbar\ \eqdef\ \langle\,d\,\rangle.
\end{equation}
A priori, $\dbar$ can be a function of time for a moving bottom. However, via a change of vertical 
coordinate $y^{\star}\eqdef y+\dbar-d_0$ ($d_0$ a constant) it is always possible to consider 
$\dbar$ independent of $t$. In that case, $g$ is a function of time $t$ and the frame of reference 
is no longer Galilean in the vertical direction. Thus, from now on, we assume that $g=g(t)$ and 
that $\dbar$ is constant.


\subsection{Lagrangian for the regularised Saint-Venant equations}

\citet{ClamondDutykh2018a} have shown that regularised Saint-Venant (rSV) equations for flat 
bottoms can be obtained from the Lagrangian density
\begin{equation}\label{eq:defLepsbar}
\overline{\mathscr{L}}_\epsilon\ \eqdef\ \half\,h\,u^2\ -\ \half\,g\,h^2\ +\,\left(\,h_t\,
+\left[\,h\,u\,\right]_x\,\right)\phi\ +\ \half\,\epsilon\,h^2\left(\,h\,u_x^{\,2}\,-\, 
g\,h_x^{\,2}\,\right),
\end{equation}
where $u(x,t)$ is the depth-averaged horizontal velocity of fluid particles and $\epsilon
\geqslant0$ is a regularisation parameter, which controls the `magnitude' of regularisation. 
In other words, one can see $\epsilon$ as a measure of the `width' of regularised shock-wave 
solutions \citep{ClamondDutykh2018a}. The resulting Euler--Lagrange equations yield the classical 
Saint-Venant equations if $\epsilon=0$ and a regularisation of the latter if $\epsilon>0$ 
\citep{ClamondDutykh2018a}.

The Lagrangian density \eqref{eq:defLepsbar} yields the correct Saint-Venant equations for 
constant depths, but fails to do so for varying bottoms. Indeed, the Euler--Lagrange 
equations for $\overline{\mathscr{L}}_0$ yield
\refstepcounter{equation}
\[
h_t\ +\,\left[\,h\,u\,\right]_x\ =\ 0, \qquad
u_t\ +\ u\,u_x\ +\ g\,h_x\ =\ 0,
\eqno{(\theequation{\,\mathit{a},\!\mathit{b}})}\label{ELL0}
\]
while the classical shallow water equations are \citep{Stoker1957}:
\refstepcounter{equation}
\[
h_t\ +\,\left[\,h\,u\,\right]_x\ =\ 0, \qquad
u_t\ +\ u\,u_x\ +\ g\,\eta_x\ =\ 0.
\eqno{(\theequation{\,\mathit{a},\!\mathit{b}})}\label{SWeq}
\]
The mass conservation (\ref{ELL0}{\it a}) and (\ref{SWeq}{\it a})  are identical. However, the 
momentum conservations (\ref{ELL0}{\it b}) and (\ref{SWeq}{\it b})  are identical in constant 
depth only, but differ when $d_x\neq0$. This discrepancy is due to the potential energy term in 
equation (\ref{eq:defLepsbar}) that is evaluated from the seabed instead of the free surface 
(i.e., the potential energy density in $\overline{\mathscr{L}}_0$ is $\half\/g\/h^2$ instead 
of $\half\/g\/\eta^2$). This is of no consequence in constant depth but it is incorrect in presence 
of an uneven bottom. Thus, this issue is addressed with the Lagrangian density for the classical 
shallow water equations
\begin{equation}\label{defL0}
\mathscr{L}_0\ \eqdef\ \half\,h\,u^2\ -\ \half\,g\,\eta^2\ +\,\left(\,h_t\,+\left[\,h\,u\,\right]_x
\,\right)\phi, 
\end{equation}
and a suitable regularisation of these equation has to be introduced. 

In \citep{ClamondDutykh2018a}, the regularised Lagrangian density (\ref{eq:defLepsbar}) is 
obtained re-injecting the momentum equation into the Lagrangian density as
\begin{equation}\label{defLepsbarprime}
\overline{\mathscr{L}}_\epsilon^{\,\prime}\ \eqdef\ \overline{\mathscr{L}}_0\ +\ \sixth\,\epsilon\,
h^3\left[\,u_t\,+\,u\,u_x\,+\,g\,h_x\,\right]_x.
\end{equation}
The Lagrangian density (\ref{defLepsbarprime}) reduces to the simplified form (\ref{eq:defLepsbar}) 
after integrating by parts the extra terms and omitting the resulting boundary terms (i.e., 
$\overline{\mathscr{L}}_\epsilon^{\,\prime}\,-\,\overline{\mathscr{L}}_\epsilon\,\equiv\,[\cdots]_t
\,+\,[\cdots]_x$, see \citep[\S3.1]{ClamondEtAl2017} for details), so $\overline{\mathscr{L}}_\epsilon
^{\,\prime}$ and $\overline{\mathscr{L}}_\epsilon$ yield the same equations because boundary terms 
do not contribute to the Euler--Lagrange equations, but $\overline{\mathscr{L}}_\epsilon$ yields 
somewhat simpler derivations. 

According to the discussion above, for varying bottoms, a regularised Lagrangian density candidate 
is  
\begin{equation}\label{defLepstildeprime}
\widetilde{\mathscr{L}}_\epsilon^{\,\prime}\ \eqdef\ \mathscr{L}_0\ +\ \sixth\,\epsilon\,h^3
\left[\,u_t\,+\, u\,u_x\,+\,g\,\eta_x\,\right]_x,
\end{equation} 
that can be easily reduced, after integrations by parts and omitting boundary terms, to the 
equivalent simplified form 
\begin{equation}\label{defLepstilde}
\widetilde{\mathscr{L}}_\epsilon\ \eqdef\ \half\,h\,u^2\ -\ \half\,g\,\eta^2\ +\, 
\left(\,h_t\, + \left[\,h\,u\,
\right]_x\,\right)\,\phi\ +\ \half\,\epsilon\,h^2\left(\,h\,u_x^{\,2}\,-\,g\,h_x\,\eta_x\,\right),
\end{equation}
following the procedure described in \citep{ClamondDutykh2018a,ClamondEtAl2017}, slightly modified to 
accommodate the varying depth. However, the Lagrangian densities (\ref{defLepstildeprime}) and  
(\ref{defLepstilde}) yield unbalanced equations, i.e., the still water ($u=\eta=0$) is not a solution 
of the equations if $d_x\neq0$ (see Appendix \ref{appunbal}). Therefore, the Lagrangian density 
(\ref{defLepstilde}) is not suitable for general varying bottoms and an alternative Lagrangian must 
then be introduced. In order to derive such a suitable Lagrangian density, we note first that the 
densities of kinetic and potential energies of (\ref{defLepstilde}) are 
\refstepcounter{equation}
\[
\widetilde{\mathscr{K}}_\epsilon\ \eqdef\ \half\,h\left(\,u^2\,+\,\epsilon\,h^2\,u_x^{\,2}
\,\right), \qquad 
\widetilde{\mathscr{V}}_\epsilon\ \eqdef\ \half\,g\left(\,\eta^2\,+\,\epsilon\,h^2\,h_x
\,\eta_x\,\right). \eqno{(\theequation{\,\mathit{a},\!\mathit{b}})}\label{defKVepstilde}
\]
The regularising term in the kinetic energy can be interpreted physically as modelling 
a vertical velocity, while mathematically it is a control of the first derivative of the 
horizontal velocity. The corresponding term in the potential energy is not a proper control 
of the free surface slope if the bottom varies; such a suitable control is obviously obtained 
substituting $\eta_x^{\,2}$ for $h_x\/\eta_x$. Thus, an obvious Lagrangian density for the  
regularised shallow water equations with varying bottom is     
\begin{align}\label{defLeps}
\mathscr{L}_\epsilon\ &\eqdef\ \half\,h\,u^2\ -\ \half\,g\,\eta^2\ +\, \left(\,h_t\, + \left[\,
h\,u\,\right]_x\,\right)\,\phi\ +\ \half\,\epsilon\,h^2\left(\,h\,u_x^{\,2}\,-\,g\,\eta_x^{\,2}
\,\right).
\end{align}

Note that the derivation of regularised Saint-Venant equations with varying bottom 
is quite easy from the variational principle. It is almost intractable to derive such 
a model by tweaking directly the equations, while preserving good properties such as 
Galilean invariance, conservation laws, well balancing, etc. We could have introduced 
the regularised equations at once and study their properties, showing afterwards that 
they have several desirable characteristics. However, we find more enlightening to 
explain where and why there are issues with the original model and how we address them. 
Note also that the regularisation above is only one possibility among (possibly) many 
others, but it is not our purpose here to derive and compare several regularisations.       

{ It should be noted that, in this paper, we always assume that $d$ is a prescribed 
(i.e., known) function of $x$ and $t$. We then use $h$ or $\eta$ indifferently in the 
equations, choosing the variable providing the clearest formulation. }


\subsection{Regularised Saint-Venant equations}

The Euler--Lagrange equations for the Lagrangian density (\ref{defLeps}) are
\begin{align}
\delta\phi: \quad 0\ &=\ h_t\ +\, \left[\,h\,u\,\right]_x, \label{eq:dLdphi2D} \\
\delta u: \quad 0\ &=\ h\,u\ -\ h\,\phi_x\ -\ \epsilon\,\left[\,h^3\,u_x\,\right]_x, \\
\delta\eta: \quad 0\ &=\ \half\,u^2\ -\ g\,\eta\ -\ \phi_t\ -\ u\,\phi_x\ +\ \threehalf\,
\epsilon\,h^2\,u_x^{\,2}\ +\ \epsilon\,g\left[\,h^2\,\eta_x\,\right]_x\ -\ \epsilon\,g\,h\,\eta_x^{\,2},
\end{align}
thence
\begin{align}
\phi_x\ &=\ u\ -\ \epsilon\,h^{-1}\left[\,h^3\,u_x\,\right]_x, \\
\phi_t\ &=\ -\,\half\,u^2\ -\ g\,\eta\ +\ \epsilon\,h^{-1}\,u\left[\,h^3\,u_x\,\right]_x\ +\ 
\threehalf\,\epsilon\,h^2\,u_x^{\,2}\ +\ \epsilon\,g\,h^2\,\eta_{xx}\nonumber\\
&\quad +\ 
\epsilon\,g\,h\,\eta_x\left(h_x+d_x\right).
\end{align}
Eliminating the variable $\phi$ between these last two relations one obtains
\begin{gather}
\partial_t\left\{\,u\,-\,\epsilon\,h^{-1}\left[\,h^3\,u_x\,\right]_x\,\right\}\,+\ 
\partial_x\left\{\,\half\,u^2\,+\,g\,\eta\,-\,\epsilon\,h^{-1}\,u\left[\,h^3\,
u_x\,\right]_x\, \right.\hspace{22mm} \nonumber\\ 
\left.\quad\, -\, \threehalf\,\epsilon\,h^2\,u_x^{\,2}\, -\, \epsilon\,g\,h^2\,\eta_{xx}\, 
-\, \epsilon\,g\,h\,\eta_x\left(h_x+d_x\right)\,\right\}\,=\ 0. \hspace{-17mm}\label{eq:momtan}
\end{gather}

Equations (\ref{eq:dLdphi2D}) and (\ref{eq:momtan}) form a regularised Saint-Venant system 
for a varying bottom. (\ref{eq:momtan}) describes the conservation of the tangential momentum 
at the free surface \citep{ClamondEtAl2017,GavrilyukEtAl2015}. Several equations for momentum 
and total energy fluxes can be subsequently derived as
\begin{equation}
u_t\ +\ u\,u_x\ +\ g\,\eta_x\ +\ \epsilon\,h^{-1}\left[\,h^2\,R\,\right]_x\ =\ 
\epsilon\,g\,h\,\eta_x\,d_{xx}, \label{mom}
\end{equation}
\begin{equation}
\left[\,h\,u\,\right]_t\ +\, \left[\,h\,u^2\, +\, \half\,g\,h^2\, +\, \epsilon\,
h^2\,R\,\right]_x\ =\ g\,h\,d_x\ +\ \epsilon\,g\,h^2\,\eta_x\,d_{xx}, \label{eq:momflu}
\end{equation}
\begin{gather}
m_t\ +\,\left[\,m\,u\,+\,\half\,g\,h^2\,-\,\epsilon\,h^2\left(\,2\,h\,u_x^{\,2}\,+\,g\,h\,
\eta_{xx}\,+\,\half\,g\,\eta_x^{\,2}\,+\,g\,\eta_x\,d_x\,\right)\,\right]_x \nonumber\\ 
=\ g\,h\,d_x\ +\ \epsilon\,g\,h^2\,\eta_x\,d_{xx}, \label{eq:momfluconj} \\
\left[\,\half\,h\,u^2\,+\,\half\,\epsilon\,h^3\,u_x^{\,2}\,+\,\half\,g\,\eta^2\,+\,\half\,
\epsilon\,g\,h^2\,\eta_x^{\,2}\,\right]_t \nonumber\\ 
+\,\left[\,\half\,h\,u^3\,+\,\half\,\epsilon\,h^3\,u\,u_x^{\,2}\,+\,g\,h\,u\,\eta\,+\,
\epsilon\,g\,h^3\,\eta_x\,u_x\,+\,
\epsilon\,h^2\,u\,R\,\right]_x\nonumber\\ 
=\ \half\,\dot{g}\left(\,\eta^2\,+\,\epsilon\,h^2\,\eta_x^{\,2}\,\right)\, -\ g\,\eta\,d_t\ 
-\ \epsilon\,g\,h^2\,\eta_x\,d_{xt}, \label{eq:enerSV}
\end{gather}
with $m\eqdef h\/u\/-\/\epsilon\left[h^{\,3}\/u_{\,x}\right]_x$, $\dot{g}\eqdef\ud g/\ud t$ 
and 
\begin{align}\label{defR}
R\ &\eqdef\ h\left(\,u_x^{\,2}\, -\, u_{xt}\, -\, u\,u_{xx}\,\right)\, -\ g\left(\,h\,\eta_{xx}\, 
+\,\half\,\eta_{x}^{\,2}\,+\,\eta_x\,d_x\,\right) \nonumber \\
  &=\ 2\,h\,u_x^{\,2}\ -\ \half\,g\left(\,h_x^{\,2}\,-\, d_x^{\,2}\,\right)\, 
  -\ h\left[\,u_t\, +\, u\,u_x\,+\,g\,\eta_x\,\right]_x.
\end{align}
Equation (\ref{eq:momfluconj}) for the momentum flux is particularly helpful in revealing 
the Hamiltonian structure of regularised Saint-Venant equations.


\subsection{Hamiltonian formulation}\label{sec:ham}

Let be the Hamiltonian functional density 
\begin{equation}\label{eq:defHeps}
\mathscr{H}_\epsilon\ \eqdef\ \half\,h\,u^2\ +\ \half\,g\,(h-d)^2\ +\ \half\,\epsilon\,
h^3\,u_x^{\,2}\ +\ \half\,\epsilon\,g\,h^2\,(h_x-d_x)^2,
\end{equation}
and the momentum $m\eqdef\mathcal{E}_u\{\mathscr{H}_\epsilon\}=h\/u-\epsilon\left[\/h^3\/u_x\/\right]_x$ 
where $\mathcal{E}_u$ is the Euler--Lagrange operator with respect of the variable $u$. The variables 
$m$ and $u$ are related via a linear non-autonomous self-adjoint positive-definite (because $h$ and 
$\epsilon$ are positive) Sturm--Liouville operator $\mathcal{L}_h\eqdef h-\epsilon\partial_x[\/h^3\/
\partial_x\/]$, i.e., $m=\mathcal{L}_h\{u\}$ that can be inverted as $u=\mathcal{G}_h\{m\}$ with 
$\mathcal{G}_h\eqdef\mathcal{L}_h^{\,-1}$. Expressing the Hamiltonian functional density 
(\ref{eq:defHeps}) as function of $h$ and $m$, we have
\begin{align}
\mathcal{E}_m\{\,\mathscr{H}_\epsilon\,\}\ &=\ \mathcal{G}_h\left\{\,h\,\mathcal{G}_h\{\,m\,\}\,
\right\}\,-\ \epsilon\,\mathcal{G}_h\,\partial_x\left\{\,h^3\,\partial_x\,\mathcal{G}_h
\{\,m\,\}\,\right\}\,=\ \mathcal{G}_h\,\mathcal{L}_h\,\mathcal{G}_h\{\,m\,\}\ =\ u, \label{eq:dH1dm} \\
\mathcal{E}_h\{\,\mathscr{H}_\epsilon\,\}\ &=\ g\,\eta\ -\ \epsilon\,g\,h\,\eta_x\left(h_x+d_x\right)\, 
-\ \epsilon\,g\,h^2\,\eta_{xx}\ -\ \half\,u^2\ -\ \threehalf\,\epsilon\,h^2\,u_x^{\,2}. \label{eq:dH1dh}
\end{align}
The derivation of \cref{eq:dH1dm} is straightforward because $\mathcal{G}_{\,h}$ is self-adjoint, 
but the derivation of \cref{eq:dH1dh} is more involved. The latter is obtained exploiting the relations
\begin{align}
\mathcal{L}_{h+\delta h}\ &=\ h\ +\ \delta\/h\ -\ \epsilon\,\partial_x\,(h+\delta\/h)^3\,
\partial_x\ =\ \mathcal{L}_h\ +\ \delta\/h\ -\ 3\,\epsilon\,\partial_x\,h^2\,\delta\/h\,
\partial_x\ +\ \mathrm{O}\!\left((\delta\/h)^2\right) \nonumber \\
&=\ \mathcal{L}_h\left[\,1\,+\,\mathcal{G}_h\,\delta\/h\, -\, 3\,\epsilon\,
\mathcal{G}_h\,\partial_x\,h^2\,\delta\/h\,\partial_x\,\right]\ +\ 
\mathrm{O}\!\left((\delta\/h)^2\right).
\end{align}
Thence, inverting this relation,
\begin{align}
\mathcal{G}_{h+\delta\/h}\ &=\,\left[\,1\,+\,\mathcal{G}_h\,\delta\/h\,-\,3\,\epsilon\,
\mathcal{G}_h\,\partial_x\,h^2\,\delta\/h\,\partial_x\,\right]^{-1}\,\mathcal{G}_h\ +\ 
\mathrm{O}\!\left((\delta\/h)^2\right) \nonumber \\
&=\ \mathcal{G}_h\ -\ \mathcal{G}_h\,\delta\/h\,\mathcal{G}_h\ +\ 3\,\epsilon\,\mathcal{G}_h
\,\partial_x\,h^2\,\delta\/h\,\partial_x\,\mathcal{G}_h\ +\ \mathrm{O}\!\left((\delta\/h)^2
\right).
\end{align}
Thus, for the kinetic energy functional $\mathfrak{K}(h,m)=\int\mathscr{K}\,\ud\/x$ 
with density\footnote{Integrating by parts, we have $h\/u^2\/+\/\epsilon\/h^3\/u_x^{\,2}\/
=\/u\/\mathcal{L}_h\{\/u\/\}\/+\/\left[\/\epsilon\/h^3\/u\/u_x\/\right]_x\/=\/ m\/
\mathcal{G}_h\{\/m\/\}\/+\/\text{`boundary\ terms'}$, so the kinetic energy part of the 
Hamiltonian density (\ref{eq:defHeps}) can be replaced by $\half\/m\/\mathcal{G}_h\{\/m\/\}$.} 
$\mathscr{K}\eqdef\half\,m\,\mathcal{G}_h\{\,m\,\}$, we obtain
\begin{align}
\mathfrak{K}(h+\delta\/h,m)\ -\ \mathfrak{K}(h,m)\ &=\ -\,\frac{1}{2}\int m\,\mathcal{G}_h
\{\,\delta\/h\,u\,\}\,\ud\/x\ +\ \frac{3\,\epsilon}{2}\int m\,\mathcal{G}_h\,\partial_x\,
\{\,h^2\,\delta\/h\,u_x\,\}\,\ud\/x \nonumber \\
&=\ -\,\frac{1}{2}\int u\,\delta\/h\,u\,\ud\/x\ -\ \frac{3\,\epsilon}{2}\int u_x\,h^2\,
\delta\/h\,u_x\,\ud\/x\,,
\end{align}
where we have exploited the self- and skew-adjointness of, respectively, $\mathcal{G}_h$ and 
$\mathcal{G}_h\,\partial_x$. It follows immediately that $\mathcal{E}_h\{\mathscr{K}\}\/=\/-\half
\/u^2\/-\/\threehalf\/\epsilon\/h^2\/u_x^{\,2}$ and the equation (\ref{eq:dH1dh}) is subsequently 
obtained at once.

Finally, the (non-canonical) Hamiltonian structure takes the form
\begin{gather}
\partial_t\binom{h}{m}\ =\ -\,\mathbb{J}\scal\binom{\mathcal{E}_h\{\mathscr{H}_\epsilon\}}
{\mathcal{E}_m\{\mathscr{H}_\epsilon\}}\ =\ -\left[
  \begin{array}{cc}
    0 & \partial_x\,h \\
    h\,\partial_x & m\,\partial_x\,+\,\partial_x\,m
  \end{array}
  \right]
\scal\binom{\mathcal{E}_h\{\mathscr{H}_\epsilon\}}{\mathcal{E}_m\{\mathscr{H}_\epsilon\}}\,= \nonumber \\
  \left(\begin{array}{c}
    {-\left[\,h\,u\,\right]_x} \\
    {g\/h\/d_x\/+\/\epsilon\/g\/h^2\/\eta_x\/d_{xx}\/-\left[\/u\/m\/+\/\half\/g\/h^2\/-\/
    \epsilon\/h^2\left(\/2\/h\/u_x^{\,2}\/+\/g\/h\/\eta_{xx}\/+\/\half\/g\/\eta_x^{\,2}\/+\/ 
    g\/\eta_x\/d_x\/\right)\,\right]_x}
  \end{array}\right),
\end{gather}
yielding the equations (\ref{eq:dLdphi2D}) and (\ref{eq:momfluconj}). It should be noted that $\mathbb{J}$ 
being skew-symmetric and satisfying the Jacobi identity \citep{Nutku1987}, it is a proper Hamiltonian 
(Lie--Poisson) operator.


\begin{figure}
  \centering\capstart
  \includegraphics[width=0.79\textwidth]{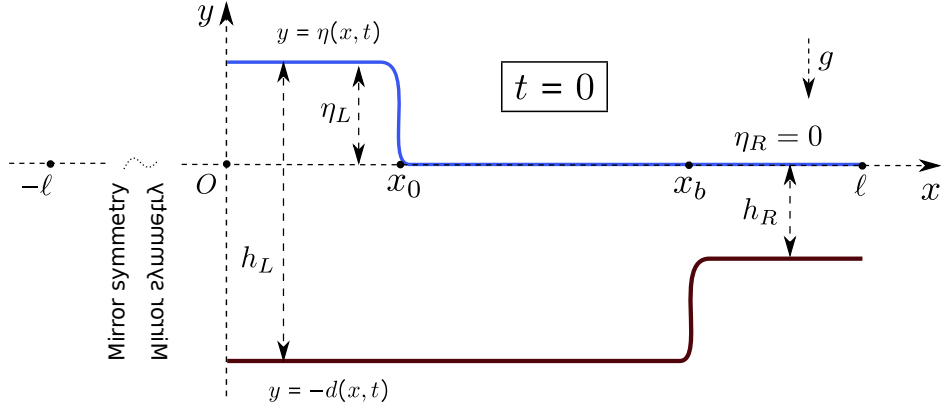}
  \caption{\em Sketch of the numerical test case considered in section \ref{sec:num}.}
  \label{fig:test}
\end{figure}

\begin{figure}
  \centering\capstart
  \includegraphics[width=0.79\textwidth,height=6cm]{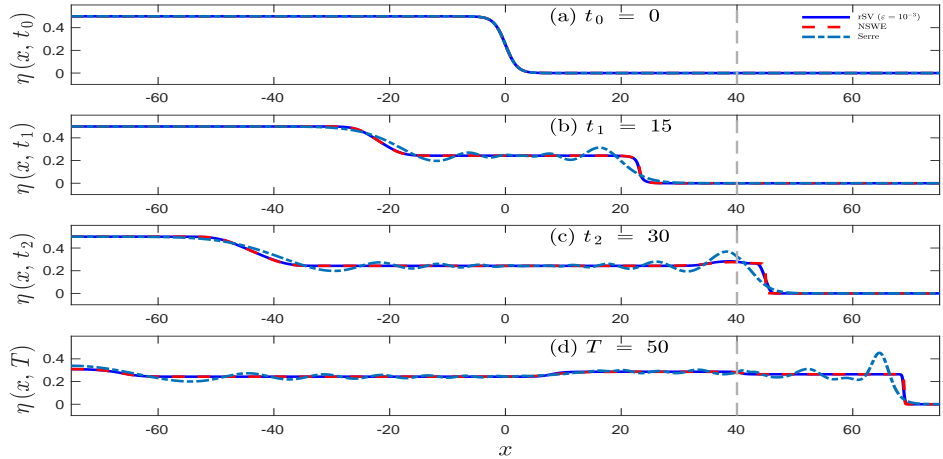}
  \caption{\em Evolution of a step initial condition under the dynamics of the regularised 
  Saint-Venant (rSV), NSWE and the Serre equations. The vertical (gray) 
  dashed line indicates the position of the bottom step.}
  \label{fig:2}
\end{figure}

\begin{figure}
  \centering\capstart
  \includegraphics[width=0.79\textwidth,height=5cm]{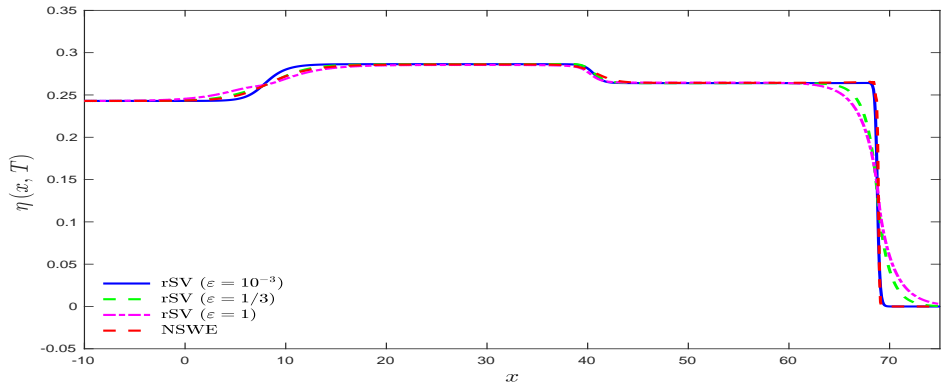}
  \caption{\em Zoom on a portion of the computational domain at the final simulation time. For the 
  sake of clarity, we report the regularised (rSV) (for three values of $\epsilon$) and classical 
  (NSWE) shallow water equations only. 
  Note the exact coincidence of the main shock position around $x \approx 70$.}
  \label{fig:3}
\end{figure}

\section{Numerical results}
\label{sec:num}

In this section, we compare the rSV, NSWE and Serre--Green--Naghdi (SGN) systems for the formation 
and propagation of shock waves and their interaction with a variable bathymetry using high order 
numerical methods. We include into our comparisons the SGN equations in order to illustrate the 
typical behaviour of a fully nonlinear but (weakly) dispersive system in the same conditions. 

We consider a periodic initial value problem (IVP) on the computational domain $x\in[-\ell;\ell]$ 
although, due to the symmetry, we shall plot only the sub-domain $x\in[0;\ell]$. The SGN system is 
solved numerically using the standard Galerkin/finite-element method with smooth cubic splines on 
an uniform grid together with a fourth-order Runge--Kutta method for the temporal discretisation, 
as described and analysed in \citep{MitsotakisEtAl2014} for a flat bottom and in \citep{MitsotakisEtAl2017} 
for varying bottoms. This method can perform really well for smooth 
solutions due to its conservative properties. 
{ When it comes to describe nearly discontinuous solutions, then spurious oscillations may appear due 
to the Gibbs phenomenon. In order to avoid this phenomenon, one can use artificial diffusion, a method that 
is commonly used for the numerical solution of hyperbolic conservation laws. Specifically, by adding a 
diffusion term $\delta u_{xx}$ we were able to control the generation of spurious oscillations at the level
where the solution remained practically unaltered. In the following experiment, we took the value of $\delta$ 
to be $0.05$. This specific value appeared to be the optimal one for this experiment, independent of the 
choice of the mesh parameter $\Delta x$.}
The rSV equations are solved by the same numerical method which was appropriately adapted to the analogous 
terms. { Note that our goal here is simply to illustrate the behaviour of the rSV equations with a 
classical numerical scheme. A non-dissipative numerical scheme exploiting the properties of the rSV equations 
is left for a future work.}
 
In all simulations the grid length for the spatial discretisation is $\Delta x = 0.1$ and the time step 
$\Delta t=0.01$. The NSWE 
equations are solved using a finite volume (FV) method (described in \citep{DutykhEtAl2011}) in the same 
interval and with the discretisation parameters used for the numerical solution of the regularised and 
dispersive systems. Namely, for the FV method we used the HLL numerical flux function and the second 
order UNO2 reconstruction with the \texttt{minmod} limiter \citep{DutykhEtAl2011}.

\begin{table}
{\small
  \centering\capstart
  \caption{\em Various physical parameters used in numerical simulations. See also \cref{fig:test} for an illustration.}
  \begin{tabular}{lc}
    \toprule
    \textit{Parameter} & \textit{Value} \\
    \midrule
    Gravity acceleration $g$ & $1$ \\
    Computational domain half-length $\ell$ & $200$ \\
    Still water depth on the left $d_L$ & $2$ \\
    Free surface elevation on the left $\eta_L$ & $0.5$ \\
    Still water depth on the right $d_R$ & $1$ \\
    Free surface elevation on the right $\eta_R$ & $0$ \\
    Initial shock wave position $x_0$ & $0$ \\
    Bottom step location $x_b$ & $40$ \\
    Final simulation time $T$ & $50$ \\
    Regularisation parameter $\eps$ & $0.001$ \\
    Transition length parameter $\delta$ & $0.5$ \\
    \bottomrule
  \end{tabular}
  \label{tab:params}
  }
\end{table}

The test case used in the present study is schematically shown in Figure \ref{fig:test}, and the values 
of various physical parameters are reported in Table \ref{tab:params}. Namely, we consider the standard 
benchmark of the dam break problem with a variable bathymetry. The bottom (a simple { steady} smooth 
step) and the initial condition for the free surface are given, respectively, by 
\begin{align}
d(x,t)\ &=\ d_L\ -\ \half\,(d_L-d_R)\left[\,1\,-\,\tanh\!\left(\delta(x-x_b)\right)\,\right], \\
\eta(x,0)\ &=\ \eta_L\ -\ \half\,(\eta_L-\eta_R)\left[\,1\,-\,\tanh\!\left(\delta(x-x_0)\right)\,\right].
\end{align}
The velocity field is taken to be initially zero, i.e., $u(x,0)=0$.

The simulation results are shown in Figure \ref{fig:2}, where we present the free surface elevation 
initially, in the middle of simulation and at the final time $t=T$ (four snapshots in total). A zoom 
of the free surface elevation at the final time is shown in Figure \ref{fig:3}. One can see an excellent 
agreement between NSWE and rSV systems. In particular, the shock positions coincide perfectly. This 
very visible on the zoomed figure \ref{fig:3}, where we have also reported three different values 
of $\epsilon$. In all cases, the position of the regularised shock (inflexion point of the 
free surface) is exactly the same as with the NSWE, as it is the case in constant depth 
\citep{ClamondDutykh2018a,PuEtAl2018}.
The absence 
of oscillations in the rSV solution confirms the absence of any dispersion, as it was designed for. 
In contrast, the weakly dispersive SGN system develops oscillations in the same experimental conditions.


\section{Conclusion and perspectives}\label{sec:cp}

In this article, we proposed a new regularisation for the nonlinear shallow water equations 
(NSWE) over general uneven bathymetries. The derivation follows a variational procedure described 
in previous works \citep{ClamondDutykh2018a,ClamondEtAl2017}. This method has the advantage of 
being automatically conservative and the resulting equations are also well-balanced,  
non-dispersive and non-dissipative. The regularised Saint-Venant (rSV) equations thus obtained 
possess several conservation laws. Moreover, the regularised system possesses also a Hamiltonian 
formulation, as the original equations do \citep{LeimkuhlerReich2005}. 
Finally, the rSV system was studied numerically with 
the finite element method (FEM). The solutions of the rSV system were compared to the classical NSWE 
(solved with FV) and the Serre--Green--Naghdi equations (solved with FEM 
as well). The numerical results confirmed the absence of dispersive effects in fully nonlinear simulations. 
An excellent agreement with NSWE could be noticed as well.

Concerning the perspectives, the generalisation of these results to 3D flows (i.e., two horizontal 
dimensions) is the next natural step in this research direction.


\subsection*{Acknowledgments}
\addcontentsline{toc}{subsection}{Acknowledgments}

D. Mitsotakis acknowledges the support by the RSL fund of Victoria University of Wellington 
and by the University Savoie Mont Blanc, LAMA UMR 5127, for the hospitality during his stay 
in March 2019.





\appendix
\section{Unbalanced equations}\label{appunbal}

The Euler--Lagrange equations for the functional density (\ref{defLepstilde})  
yield, after some algebra, the mass conservation $h_t+\left[\/h\/u\/\right]_x\/ =\/ 0$ 
and the conservation of momentum 
\begin{align}
\partial_t\left\{\,u\,-\,\epsilon\,h^{-1}\left[\,h^3\,u_x\,\right]_x\,\right\}\, +\ 
\partial_x\left\{\,\half\,u^2\, +\, g\,\eta\,-\,\epsilon\,h^{-1}\,u\left[\,h^3\,u_x\,
\right]_x\  \right.\ & \nonumber \\
\left.  -\,\epsilon\,h^2\left(\,\threehalf\,u_x^{\,2}\, +\, g\,h_{xx}\, -\, \half\,g\,
d_{xx}\,\right) -\, \epsilon\,g\,h\,h_x^{\,2}\,\right\}\,&=\ 0. \label{momunbal}
\end{align}
For still water --- i.e., when $u=\eta=0$ and $h=d(x)$ --- the mass conservation is 
satisfied identically and the momentum conservation (\ref{momunbal}) becomes, after 
simplifications, 
\begin{align}
-\/\sixth\,\epsilon\,g\left[\,d^3\,\right]_{xxx}\ =\ 0. \label{momunbal0}
\end{align}
Thus, the stil water is solution if $\epsilon=0$ (classical shallow water equations) 
and, when $\epsilon>0$, if $d^3$ is a second-order polynomial in $x$. For general bottoms, 
the equations derived from (\ref{defLepstilde}) are not well-balanced 
and, therefore, the Lagrangian density (\ref{defLepstilde}) does not provide a suitable 
regularisation of the classical shallow water equations for uneven bottoms. 


\end{document}